\documentclass{article}
\usepackage{times}
\usepackage{amsmath}
\usepackage{amssymb}
\usepackage{array}
\usepackage{epsfig}
\usepackage{pstricks}
\newcommand{\inc}{^{\prime\prime}}

\title{ From Baking a Cake to Solving the Schr\"odinger  Equation } 
\author{
Edward A. Olszewski  \\ 
 {\em Department of Physics} \\ 
 {\em University of North Carolina at Wilmington} \\  
 {\em Wilmington, North Carolina 28403-5606} \\
{\em email: olszewski@uncw.edu}
}
\date{}

\begin{document}

\bibliographystyle{plain}

\maketitle

\begin{abstract}
The primary emphasis of this study has been to explain how modifying a cake recipe by changing either the dimensions of the cake or the amount of cake batter alters the baking time.  Restricting our consideration to  the g\'{e}noise, one of the basic cakes of classic French cuisine, we have obtained a semi-empirical formula for its baking time as a function of oven temperature, initial temperature of the cake batter, and dimensions of the unbaked cake.  The formula, which is based on the Diffusion equation, has three adjustable parameters whose values are estimated from data obtained by baking g\'{e}noises in cylindrical pans of various diameters.  The resulting formula for the baking time exhibits the scaling behavior typical of diffusion processes, i.e.\@ the baking time is proportional to the $\text{ (characteristic length scale)}^2$ of the cake. It also takes account of evaporation of moisture at the top surface of the cake, which appears to be a dominant factor affecting the baking time of a cake.  In solving this problem we have obtained solutions of the Diffusion equation which are interpreted naturally and straightforwardly in the context of heat transfer; however, when interpreted in the context of the Schr\"{o}dinger equation, they are somewhat peculiar.  The solutions describe a system whose mass assumes different values in two different regions of space.  Furthermore, the solutions exhibit characteristics similar to the evanescent modes associated with light waves propagating in a wave guide.  When we consider the Schr\"{o}dinger equation as a non-relativistic limit of the Klein-Gordon equation so that it includes a mass term, these are no longer solutions.   

\end{abstract}

\setcounter{page}{1}
\setcounter{section}{0}
\setcounter{equation}{0}
\section{Introduction}
\label{sect1}
It is an aphorism (at least to some physicists) that all {\em natural} phenomena can be understood and clarified in accord with general physical principles.  The examples of such, that come immediately to mind are those related directly to the natural sciences. There are, however, numerous other examples, where the connection to physics is blatantly obvious and tacitly assumed, but is, nonetheless, usually ignored. One notable example is the culinary arts.  In order to cook and do it well, in all likelihood, requires little knowledge of the sciences in general and physics in particular.  Indeed, most cook books are just that {\em cook books}: one merely follows the steps in a recipe, and one obtains the finished product.  In general, no information is given detailing the physical principles on which a recipe is based, making it difficult, in many instances, to modify the recipe in non-trivial ways. 

      Surprisingly, there appears to be a dearth of research related to the cake baking process. In fact, much of what appears in the literature is specifically related to bread baking. A notable exception is the work of Lostie and others~\cite{lostie02a},\cite{lostie02b}.  In that study they have analyzed, in detail, the baking process of sponge batter during the first baking period.\footnote{The baking process involves several stages before the cake is completely baked.  Their study only applies to the initial stage of the baking process.}$^,$\footnote{In reference~\cite{lostie02b} a comprehensive review of the existing research is given.}  As described in reference~\cite{lostie02a} the data used in their analysis have been obtained from highly  controlled experiments in which the cake batter has been heated only from the top, with the sides and bottom of the pan being thermally insulated.  Based on these experiments they have proposed a one dimensional model of the baking process to predict the spatial dependence and time evolution of water content (water and vapor), temperature, gas phase pressure and porosity (the proportion of non-solid volume to total volume) of the cake batter.  The proposed model has nine adjustable parameters whose values are determined by fitting the model to the data described in part I of their study.  This study appears to be the most comprehensive in providing a complete understanding of the baking of sponge batter.
  
The motivation for this article is to provide an answer, albeit a limited one, to the question of how the baking time of a cake varies when its recipe is modified, either by changing the dimensions of the cake or the amount of cake batter. In order to make the problem manageable we have restricted our analysis to one type of cake only, the g\'{e}noise, one of the basic cakes of classic French cuisine.  Although the technique for baking this cake is somewhat exacting, the ingredients should be readily available,  facilitating attempts to reproduce our results.  The theoretical basis of our solution is the Diffusion equation, and because of its intimate relation to Schr\"{o}dinger equation, its solutions, in a certain sense, can be considered dual to solutions of the Schr\"{o}dinger equation. We explore the dual interpretation of the solution to the cake baking problem and find that its interpretation from the perspective of the Schr\"{o}dinger equation to be somewhat surprising and unsual.

The outline of the article is as follows.  First, we briefly discuss the quantitative aspects of cake baking. Next, we describe the experimental techniques and data  used in the subsequent analysis.   We, then, propose a simple model for the cake baking process and use it to obtain a semiempirical formula for estimating the baking time of the g\'{e}noise. Finally, we interpret the model in the context of the Schr\"{o}dinger equation and discuss the implications.

\setcounter{equation}{0}
\section{Cake Baking as a Diffusion Process}
\label{sect2}
\subsection{Cake Baking from a Quantitative Point of View}
\label{sect2_1}
Before modeling the cake baking process, mathematically, we must first make more precise the imprecise measuring techniques that have, customarily, accompanied the instructions for baking a cake. A typical cake recipe lists the ingredients with amounts, a description of the technique used to prepare the cake batter, the baking time for the recipe, and usually some qualitative measure of determining whether the cake is baked.  Until recently, amounts of ingredients have generally been given in units of liquid measure, like the cup or tablespoon.  One can only surmise that the reason for this has been the scarcity of mass measuring scales in the home. The obvious reason for using dry measure (measuring by scale), rather than liquid measure,  is to increase the likelihood of consistently producing a cake with the desired taste, texture, and appearance. Generally, the time necessary to bake a cake is accompanied by some subjective measure of determining whether the cake has baked to the proper degree of doneness. Such subjective measures include, {\em when a toothpick thrust into the center of the cake comes out clean}, {\em when the cake shows a faint line of shrinkage from the sides of the pan}, {\em when the top springs back lightly}, {\em when you can smell the cake}, etc. To a first approximation this criterion of doneness can be quantified by equating the degree of doneness to the temperature at the center of the cake, although one may imagine more complex criteria, which include other measurable quantities, like the temperature gradient.

\subsection{The G\'enoise}
\label{sect2_2}
The cake used in this study is the g\'enoise, a basic cakes of classic French cuisine. The recipe\footnote{The only significant modification to the recipe is that the eggs have  not been warmed over boiling water prior to mixing.}  is adapted from that described in a book of Jacques P\'epin~\cite{pepin1}.\footnote{Barham provides an excellent discussion about the physics and chemistry of baking sponge cakes, of which the g\'enoise is an example~\cite{barham01}. } As is typical, amounts of most ingredients are given in liquid measure. In Table~\ref{table1}, we have converted these to dry measure.  Qualitatively, we have found that the recipe prepared   using traditional measuring techniques produces a cake consistent in texture, taste, and appearance with one prepared using dry measuring techniques.  In fact, quantitatively, amounts  measured and re-measured using traditional techniques  differ, typically, less than  two percent when compared to amounts measured by weight. 
\begin{table}
\center 
\begin{tabular} {|  c |  c  c |   } \hline
  & Traditional Measure & Dry Measure (gm) \\ \hline 
 eggs  & 6 large&  298   \\ \hline
 sugar  & 3/4 cup&  176   \\ \hline
 vanilla extract  & 1/2 tsp&  2   \\ \hline
  all purpose flour & 1 cup &   144  \\ \hline
  butter & 3/4 stick &  114   \\ \hline
\end{tabular}
\caption{The proportion of ingredients for the g\'enoise is given in traditional measure as well as dry measure. }
\label{table1}
\end{table}
According to the recipe  the cake batter is placed in two cake pans, $8^{\prime\prime}$ in diameter by $1\frac{1}{2}^{\prime\prime}$ deep, filling each pan $\frac{3}{4}$ full.  The cakes are then baked in an oven of temperature $T_o = 350^\circ$~F for a duration of  time between 22 and 25 minutes. In order to establish benchmarks for this study we have prepared the recipe, with exceptions as noted, filling two $8^{\prime\prime}$ cake pans to a depth of $1^{\prime\prime}$. Assessing the degree of doneness using traditional techniques we have observed  the  baking time to be, approximately, 17 minutes.  The qualitative measure of doneness corresponds, quantitatively, to a temperature  $T_f$ ($ T_f =203^\circ$~F) at the center of the cake.  The  recipe has been prepared several times with baking times varying by no more than two minutes.

\subsection{Theory from a Naive Perspective}
\label{sect2_3}
Initially, we  were believing that we could model the cake baking process, naively, as a simple diffusion process.\footnote{  Klamkin has considered such a model for estimating the scaling behavior in the cooking time of a roast~\cite{klamkin61}. }  Although the model is grossly inadequate, it has provided the basis of the final model and is, thus, presented.  This model derives from a solution of the Diffusion equation~\cite{kittel1}, 
\begin{equation}
 \mathcal{D} \nabla^{2} T =  \frac{\partial T}{\partial t}  \: ,  \label{eq1}
\end{equation}   
where $T = T(t, r, \theta, z)$ is the temperature of the cake at time, $t$, at location, $(r, \theta, z)$ (expressed in cylindrical coordinates), within the cake, and $ \mathcal{D}$ is the diffusion coefficient (assumed constant) for the cake batter.
We assume the cake batter to be in a cylindrical pan of radius, $R$, and thickness, $Z$,  at initial temperature $T_{c}$, and baked in an oven of constant temperature, $T_{o}$.  Assuming these initial conditions we solve the Diffusion equation for the temperature, $T$, which because of azimuthal symmetry is independent of $\theta$,
\begin{equation}
T(t,r,z)  = T_o +(T_c-T_o) \frac{2 R \sqrt{Z}}{\pi}     
          \sum_{n=1, m=1 }^\infty  \frac{[1-(-1)^n]}{n x_{m,0}}
           \Phi_{mn}     \: ,   \label{eq2}
\end{equation}
where $x_{m,0}$ is the m-th root of the zeroth order Bessel function, and   
the functions $\Phi_{mn}$ are the  normalized eigenfunctions of the Diffusion Equation, 
\begin{equation}
\begin{split}
\Phi_{mn}  =  \frac{2}{R J_1(x_{m,0}) \sqrt{Z} } & \sin(\frac{n\pi z}{Z})   
        \frac{ J_0( \frac{x_{m,0} r}{R})}{x_{m,0} J_1(x_{m,0})} \times \\
     & \times \exp[  -\{(\frac{n\pi}{Z})^2 +(\frac{x_{m,0}}{R})^2)\} \mathcal{D}t] \: . \label{eq2a}
\end{split}
\end{equation}
In Equation~\ref{eq2} the first term to the right of the equality is the steady state solution, i.e.\@ a solution of the Laplace equation.  The second term is the solution particular to the boundary conditions, i.e.\@  vanishing at the surface bounding the cylinder.  
Only the Bessel functions of zeroth order appear because of azimuthal symmetry. The other factors are required to reproduce the initial temperature on the interior of the cake at $t=0$.

Typically, the initial temperature of the cake batter $T_c \approx  80^\circ F $, corresponding to the temperature at which the cake batter is prepared. Thus, the  initial and final temperatures of the cake and temperature of the oven satisfy the relationship
\begin{equation}
\frac{T_{f}-T_{c}}{T_{o}-T_{c}} \lessapprox 1 \: . \label{eq3}
\end{equation}
For temperatures, $T_f$ which satisfy Eq.~\ref{eq3},  the infinite series, Eq~\ref{eq2}, evaluated at the center of the cake,  can be approximated by a single term 
\begin{equation}
T(t_f,0,Z/2) =T_f  = T_o +(T_c-T_o)F  \: ,     \label{eq4} 
\end{equation}
where

\begin{equation}
F= 
        \begin{cases}
\frac{8}{\pi x_{1,0} J_1(x_{1,0})}  
\exp[  -\{(\frac{\pi}{Z})^2 +(\frac{x_{1,0}}{R})^2)\} \mathcal{D}t_f]  
&  \\    \text{\hspace{.5in} if $C_1 \leq 1$ and $C_2 \leq 1$} \\ \hspace{.5in} \\
               \frac{4}{\pi} \exp[  -\{(\frac{\pi}{Z})^2\} \mathcal{D}t_f]  
& \\ \text{\hspace{.5in} if $C_1>1$}  \\ \hspace{.5in} \\
                 \frac{2}{ x_{1,0} J_1(x_{1,0})}  
\exp[  -\{((\frac{x_{1,0}}{R})^2)\} \mathcal{D}t_f]    
& \\  \text{\hspace{.5in}  if $C_2 >1$} \\ \hspace{.5in} \\
        \end{cases}
\: . \label{eq5}
\end{equation}

Here 
 \begin{equation}
C_1=\frac{2}{ x_{1,0} J_1(x_{1,0})} \exp[  -\{(\frac{x_{1,0}}{R})^2)\} \mathcal{D}t_f] \: , \label{eq6}
\end{equation}
\begin{equation} 
C_2=\frac{4}{\pi} \exp[  -\{(\frac{\pi}{Z})^2\} \mathcal{D}t_f] \: ,   \tag{\ref{eq6}$'$}
\end{equation}
and $t_f$ is the amount of  time for the temperature of the center of the cake to increase to temperature $T_f$. 
This simplification to Eq.~\ref{eq2} is most readily verified by direct calculation.

We can solve Equation~\ref{eq4} explicitly for the baking time, $t_f$, 
\begin{equation}
\frac{1}{t_f}  \propto
\begin{cases}
\frac{x_{1,0}^{2}}{R^{2}} + \frac{\pi^{2}}{Z^{2}}  
&  \\    \text{\hspace{.5in} if $C_1 \leq 1$ and $C_2 \leq 1$}  \\  \hspace{.5in} \\
               \frac{\pi^{2}}{Z^{2}}  
& \\ \text{\hspace{.5in} if $C_1>1$}  \\ \hspace{.5in} \\  
\frac{x_{1,0}^{2}}{R^{2}}    
& \\  \text{\hspace{.5in}  if $C_2 >1$} \\ \hspace{.5in} \\
        \end{cases}
\: , \label{eq7}
\end{equation}
where
\begin{equation}
\begin{split}
 x_{1,0} & \equiv \text{the first root of the zeroth order Bessel Function} \\
& \approx 2.40  
\end{split}
\: .    \label{eq8}
\end{equation}
The constant of proportionality depends only on the unknown diffusion constant, $\mathcal{D}$, which can, in principle, be estimated from the baking time of a given recipe.   In any case, if the baking time is known for a cake of specified dimensions, then Eq.~\ref{eq7} can be used to estimate the baking time of a cake of arbitrary dimensions.  For practical applications it is useful to express Eq.~\ref{eq7} with  its numerical constants appearing explicitly, in which case we have
\begin{equation}
 \frac{1}{t_f} \propto 
\begin{cases}
\frac{2.34}{D^{2}} + \frac{1}{Z^{2}}
&  \\    \text{\hspace{.5in} if $C_1 \leq 1$ and $C_2 \leq 1$} \\ \hspace{.5in} \\
\frac{1}{Z^{2}}
 & \\ \text{\hspace{.5in} if $C_1>1$}  \\ \hspace{.5in} \\  
\frac{2.34}{D^{2}}                  
& \\  \text{\hspace{.5in}  if $C_2 >1$} \\ \hspace{.5in} \\
\end{cases}
 \: , \label{eq8b}
\end{equation}
where 
$ D \equiv \text{the diameter of the cake} \: .\footnote{It is interesting to note that if the cake is rectangular, the term $2.34/D^2$ is replaced by $1/X^2+1/Y^2$, where X and Y are the length and width of the cake pan.}^{,} \footnote{The reader should be aware that  some formulae are expressed in terms of diameter rather than radius. The reason is to make transparent the relationship between these formula those applicable to rectangular cake pans. } $

 We use Eq.~\ref{eq8b} to estimate the baking time of a cake which is $2.1\inc$ radius and $4.0\inc$ depth. The calculation assumes the same initial conditions as apply in the preparation of the basic recipe described previously.\footnote{It is worth mentioning that standard cake pans are not readily available with these dimensions. This has necessitated substituting a coffee can, filled with batter to a depth of $4.0\inc$, for a cake pan.    In fact, g\'{e}noises are commonly not baked  greater than $2\inc$ in depth.} The predicted baking time is approximately 85 minutes.  The actual baking time, based on an internal temperature of $203^\circ$ F, is 26 minutes. 

\setcounter{equation}{0}

\section{Experiment}
\label{sect3}
\subsection{Procedure}
\label{sect3_1}
\begin{figure}[ht]
\begin{center}
\epsfig{file=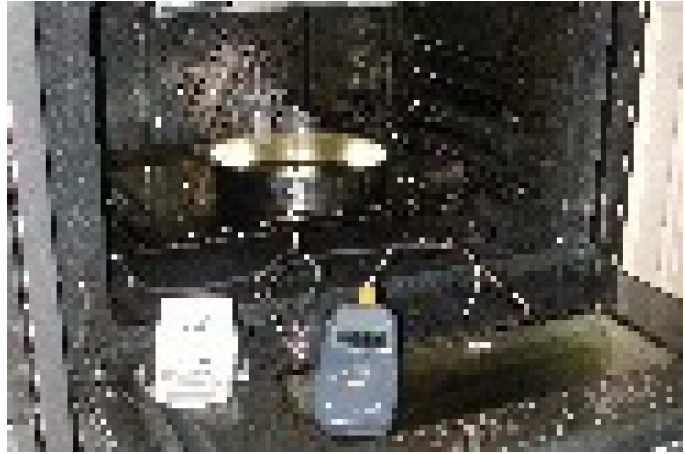,height=6cm} 
\end{center}
\caption{Shown are the oven, the baking pan, and two thermocouple sensors and temperature measuring instruments. One sensor monitors the internal temperature of the cake, and the other the temperature of the oven.}
\label{fig1}
\end{figure}

In order to understand why the previously described model of the cake baking process has failed, we have collected a variety of data for different amounts of cake batter baked in cylindrical cake pans of various dimensions.  Seven cakes have been baked whose dimensions range  from $2.1\inc$ to $6.5\inc$ radius and from $1.0\inc$ to $4.0\inc$ depth and whose baking times range from 17 to 42 minutes.  The data consist of the mass of each cake before and after baking, the diameter and depth of the cake, the depth of the cake after baking, and the temperature at the center of the cake recorded at one minute intervals until the cake is baked. The experimental setup for recording the internal temperature of the cake is shown in Figure~\ref{fig1}. 
  In most ovens the temperature of the oven is measured by a temperature probe which is close to the oven walls, and consequently, relatively distant from whatever is being baked in the oven.  The  temperature at the oven walls is typically higher than the temperature at the surface of what is being baked.  In this study temperatures have been monitored next to the cake pans, with the heating element of the oven regulated manually to maintain the desired oven temperature.  This has been done to minimize convective effects so that transfer of heat to the  cake  can be approximated as a process involving only conduction. 

The experimental procedure involves the following:
\begin{enumerate}
\item measure the mass and diameter of the cake pan;
\item prepare the cake batter;
\item fill the cake pan to the desired depth;
\item measure the mass of the cake batter;
\item place the cake in the preheated oven and record its internal temperature at one minute  intervals until the cake is baked, i.e. until the cake's internal temperature reaches $203^\circ$. \footnote{Temperature measuring instruments are located outside of the oven, so that temperature data can be collected without opening the oven door. }
\item after the cake is baked, re-measure its depth and mass.
\end{enumerate}

It is in order to make some comments about the experiment.  The baking time of individual cakes is extremely sensitive to the depth of the cake batter. The reason is that the batter is relatively shallow,  as is typical when baking cakes of this type. For example, if the cake batter is approximately $1\frac{1}{2}\inc$ deep, an error in measurement of $\frac{1}{8}\inc$ is a $12\%$ error.  Even when using extreme care to make depth measurements at the center of the cake it is difficult to insure that the depth remains constant across the surface of the cake.  This problem is exacerbated in the cakes of larger diameter used in this study.  The fact that cakes rise while baking, in itself, is not a problem, but identical cake recipes baked on different occasions do not rise precisely the same way. These uncertainties help to clarify why for many cake recipes   a range of baking times  are given whose extremes differ by $15\%$ or  more. It seems reasonable to assume that   this type of random behavior is more exaggerated in cakes of larger radius resulting in greater uncertainties in their baking times.  There is another practical consideration which is a potential source of error in  the experiment. Because of the shallow depth of the batter relative to the size of the temperature probe it is difficult to insure that  the  probe is inserted accurately into the center of the cake. On the other hand, if extremely controlled experiments are needed for making accurate theoretical predictions, then from a practical perspective the results would be of little use. One final comment, 
as it has  been impractical to automate the data collection process, specifically the temperature/time measurements, the actual data collection has been tedious.

\begin{figure}[h]
\begin{center}
\epsfig{file=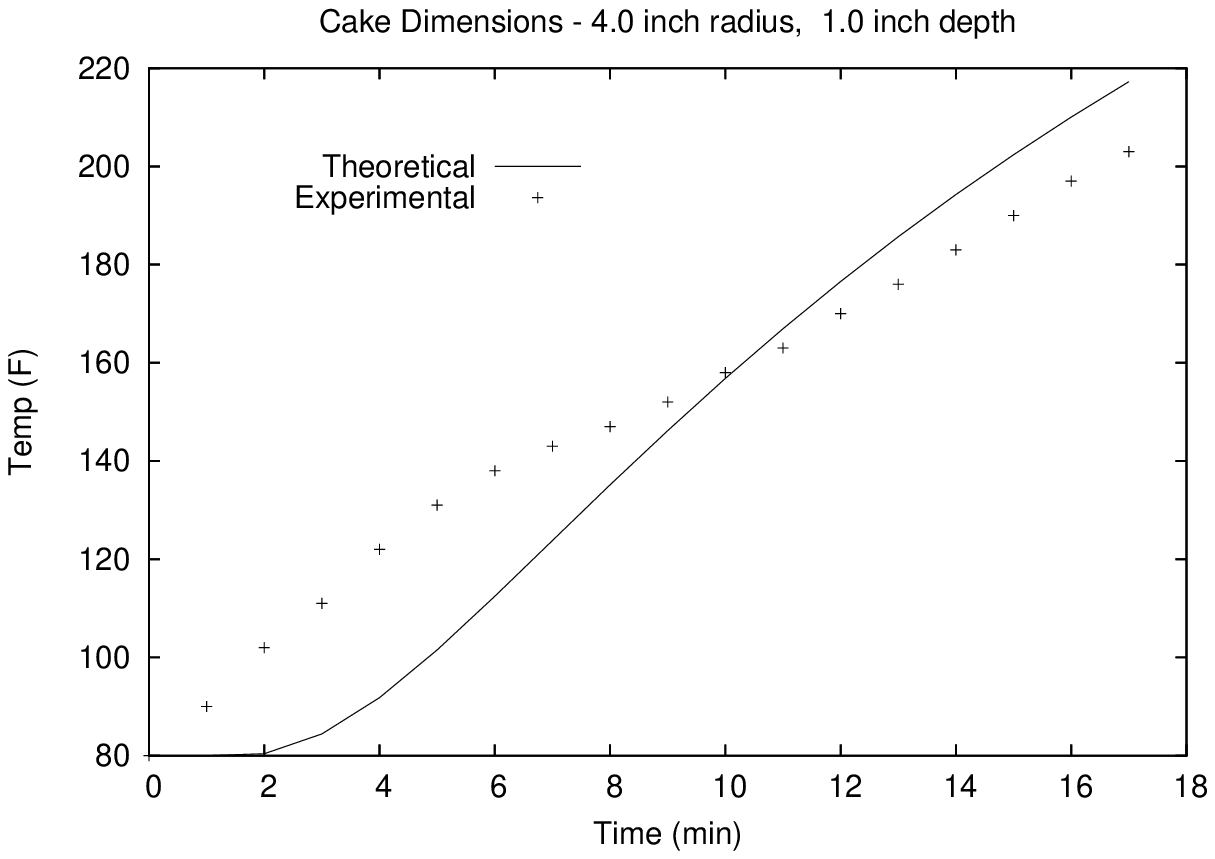,height=6cm} 
\end{center}
\caption{  The temperature of a g\'enoise $4.0\inc$ radius and $1.0\inc$ depth recorded at one minute intervals is shown along with the theoretical fit based on Eq.~\ref{eq2}.}
\label{fig2}
\end{figure}

\begin{figure}
\begin{center}
\epsfig{file=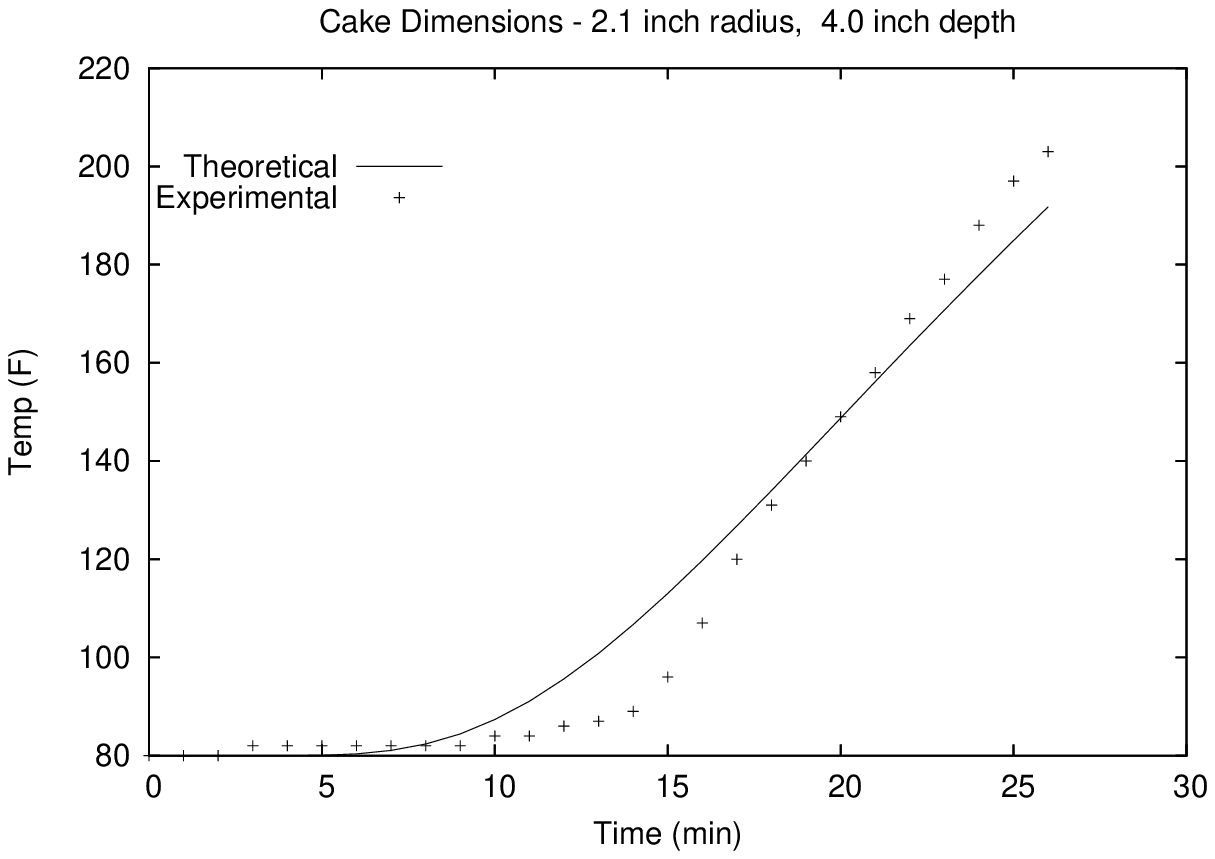,height=6cm} 
\end{center}
\caption{ The temperature of a g\'enoise $2.1\inc$  radius and $4.0\inc$ depth recorded at one minute intervals along with the theoretical fit based on Eq.~\ref{eq2}.}
\label{fig3}
\end{figure}

In Figure~\ref{fig2} we show a plot of  temperature as a function of time at the center of a cake $4.0\inc$  radius and $1.0\inc$ depth.  In addition, we also show the theoretical fit to the data based on Eq.~\ref{eq2}.  The summations in Eq.~\ref{eq2} have been truncated at $n,m=17$.\footnote{It has been necessary to truncate the sum at such  large values of $m$ and $n$ to obtain reasonable estimates of temperatures close to the initial temperature of the cake. } It appears that the model based on the Diffusion Equation is inadequate in accounting for the features of the data.
In Figure~\ref{fig3} we show a plot of  temperature as a function of time of a cake $2.1\inc$ radius and $4.0\inc$ depth.  In this case, also, the theoretical curve appears to fit the data inadequately, suggesting that a drastically different theoretical model from that proposed, initially, may be required to estimate baking times.

\subsection{Revised Model}
\label{sect3_3}
After analyzing the data both qualitatively and quantitatively, we have made several observations and drawn a number of conclusions.  We have observed that for a baked cake, not only does the volume of the cake increase, but also  the mass of the cake decreases.\footnote{We have observed that typically the mass of a baked cake is less by approximately 10\%.}  It  also seems apparent that the model based on Eq.~\ref{eq2}, exclusively, provides a poor fit to the data of temperature verses time for the cakes of various dimensions.  With hindsight it has been na\"{i}ve to believe that a model based, only, on Eq.~\ref{eq2} can account for the features of the data.  The cake batter is a viscous liquid  at the start of the baking process and a solid at the end.  Thus, not only conduction, but also  convection, and even radiation contribute to the baking process, each to varying degrees.  In addition, the parameters associated with these methods of heat transfer, thermal conductivity, density, specific heat, the diffusion coefficient, and convection coefficient, to some extent, depend on temperature.  In any case, assuming that the diffusion coefficient is constant, as in Eq~\ref{eq2}, is a simplistic assumption.  To improve the model we make the following assumptions and revise the model accordingly: 
\begin{enumerate}
\item conduction is the primary mechanism of heat transfer;
\item the thermal conductivity of the cake batter varies during the baking process, primarily, because of moisture evaporation at the top, cake surface.  
\end{enumerate}

\begin{figure}
\begin{center}
\begin{pspicture}(-2,2)(10,6)
\psellipse[fillcolor=lightgray,linecolor=black](4.0,4.0)(3.0,.25)
\psellipse[fillcolor=lightgray,fillstyle=solid](4.0,3.0)(3.0,.25)
\psframe[linecolor=lightgray,fillcolor=lightgray,fillstyle=solid](1,3)(7,5)
\psellipse[fillcolor=lightgray,fillstyle=solid,linecolor=black](4.0,5.0)(3.0,.25)
\qline(1,3.0)(1,5.0)
\qline(7,3.0)(7,5.0)
\psline{->}(.5,5)(.5,4.1)
\psline{->}(.5,2)(.5,2.9)
\psline{->}(8.5,4)(8.5,5)
\psline(0,4)(.9,4)
\psline(0,3)(.9,3)
\uput[0](.25,3.5){$a$}
\uput[0](7.5,3.5){Moisture Flow}
\uput[0] (3.7,4.4) {$\mathcal{D}^\prime$} 
\uput[0] (3.7,3.1) {$\mathcal{D}$}
\end{pspicture}
\end{center}
\caption{As heat flows into the cake, moisture evaporates through the top surface of the cake resulting in a two component system, one with a diffusion coefficient $\mathcal{D}$, corresponding to the moist batter in the bottom portion of  the pan and another with diffusion coefficient $\mathcal{D}^\prime$, corresponding to the dry batter in the top portion of the pan.}
\label{fig4}
\end{figure}
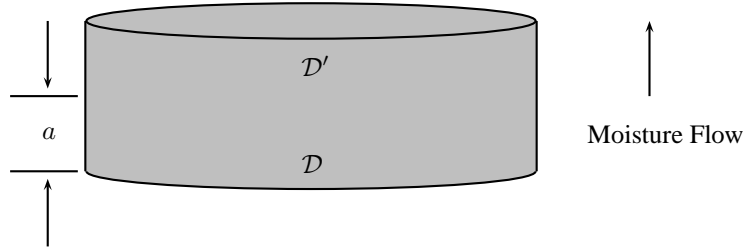

 In the revised model moisture evaporates at the top surface of the cake causing the cake batter to dry out so that, consequently, the diffusion coefficient decreases in value. The cake batter consists of two homogeneous mixtures, each with its own diffusion coefficient.  In Figure~\ref{fig4} we diagrammatically represent the process.  As baking occurs the interface  between the two homogeneous mixtures of cake batter (dry on top and moist on the bottom) moves from the top  to the bottom of the pan. The quantity $a$ denotes the interface between the two mixtures at some arbitrary time.  Here, $\mathcal{D}$ and $\mathcal{D}^\prime$ are the diffusion constants of the dry and moist mixtures, respectively. In order to reduce the time for numerical computations we have implemented the following simplification in applying the model in practice.  We have assumed that the cake batter remains uniformly moist during the time interval $0 \leq t \leq t_1$ (Interval I). Then, during the time interval $t_1 < t \leq t_2$ (Interval II) the cake comprises two homogenous mixtures,  dry batter  in the top half and  moist batter  in the bottom half of the pan, i.e.\@ the interface remains fixed at  $a=Z/2$.  At time $t_2$ and until it completes baking at time $t_{\text{f}}$, i.e.\@ $t_2 < t \leq t_{\text{f}}$ (Interval III), the cake is uniformly dry.\footnote{Any time dependence of the diffusion coefficient is assumed to result from that of the thermal conductivity, only.  Thus, we neglect possible time dependence in the mass density or specific heat. This simplifies the analysis and avoids physically unrealistic, temporal discontinuities in the temperature which result from  discontinuous changes in the diffusion coefficient.}  
 
For time $ t \leq t_1 $ and $ t_2 < t \leq  t_{\text{f}}$ the time evolution of the temperature is calculated using the  eigenfunctions of the Diffusion equation given in Eq.~\ref{eq2}, substituting  the diffusion coefficient $\mathcal{D}$ or  $\mathcal{D}^\prime$, as appropriate.  For time $t$, $ t_1 < t \leq t_2$, however, we require eigenfunctions obtained from solving the Diffusion equation for two contiguous, homogenous media whose interface is located at $z=a$.  

The unnormalized eigenfunctions, $\Psi_{mj}$, of eigenvalue, $E_{mj}$, are given by

\begin{equation}
  \Psi_{mj}= \exp(-E_{mj} t) \times J_0\left( \frac{x_{m,0} r}{R}\right) \mathcal{Z}_{mj}(z) \: . \label{eq9}
\end{equation}

Here
\begin{equation}
  \mathcal{Z}_{mj}(z)=  
        \begin{cases}
               A_{mj}  \sin(\lambda_{mj} z) (z) & 0 \leq z \leq a \\
               B_{mj}  \sin(\lambda^\prime_{mj} (Z-z))  & a < z \leq Z
        \end{cases}  
 \: .    \label{eq10}
\end{equation}
The coefficients $A_{mj}$ and $B_{mj}$ are determined by the normalization condition and the boundary conditions at $z=a$.  The quantities $\lambda_{mj} $ and $\lambda^\prime_{mj} $ are related by
\begin{equation}
  E_{mj} =\mathcal{D}\left(\lambda^2_{mj} + \left( \frac{x_{m,0}}{R}\right)^2\right) = \mathcal{D}^\prime \left(
\lambda^{\prime 2}_{mj} + \left( \frac{x_{m,0}}{R}\right)^2 \right) \: , \label{eq11}
\end{equation}
and satisfy the eigenvalue equation
\begin{equation}
   \frac{\tan(\lambda_{mj} a)}{\mathcal{D} \lambda_{mj}} = - \frac{\tan(\lambda^\prime_{mj} (Z-a))}{\mathcal{D}^\prime \lambda^\prime_{mj} } \: .   \label{eq12}
\end{equation}

The eigenvalue equation is obtained from the boundary conditions at the interface separating the two homogeneous mixtures of cake batter:
\begin{equation}
  \mathcal{Z}(z) |_{z=a^-}  = \mathcal{Z}^\prime(z) |_{z=a^+}  \: ,  \label{eq13}
\end{equation}
and
\begin{equation}
   \mathcal{D} \frac{\partial \mathcal{Z}(z)}{\partial z} \Bigg\vert_{z=a^-}  =  \mathcal{D}^\prime  \frac{ \partial \mathcal{Z}^\prime(z)}{\partial z} \Bigg\vert _{z=a^+} \: .  \label{eq14}
\end{equation}
The latter equation is the result of applying the continuity equation at the interface.
Here $\mathcal{D}>\mathcal{D}^{\prime}$ because the moist cake batter thermally conducts better than the dry cake batter.  
Expressed as a function of $\lambda^\prime_{mj}$,
\begin{equation}  \lambda^2_{mj}  = \frac{\mathcal{D}^\prime}{\mathcal{D}} \lambda^{\prime 2}_{mj}  - \left( \frac{\mathcal{D}- \mathcal{D}^\prime }{\mathcal{D}}\right) \left(\frac{x_{m,0}}{R}\right)^2      \: . \label{eq15} \end{equation}
The solutions of the eigenvalue equation exhibit the following properties:
\begin{enumerate}
\item  $ \lambda^2_{mj}  \leq \lambda^{\prime 2}_{mj}  $;
\item there are no solutions for $ \lambda^{\prime 2}_{mj} <0 \: ; $
\item the derivative of an eigenfunction is discontinuous at $z=a$.
\item Some eigenfunctions are oscillatory in the region $0 \leq z \leq a$ and decaying in the region $a < z \leq Z$.
\end{enumerate}
 Some of these properties have interesting consequences, especially when interpreted in the context of the Schr\"{o}dinger equation. We defer further discussion of these until Section~\ref{sect5}.

We now prescribe  the recipe  for implementing the revised model in practice. \begin{enumerate}
\item Specify values for the adjustable parameters $\mathcal{D}$, $\mathcal{D}^{\prime}$, $t_1$, and $t_2$.  
\item Substitute  the temperature of the oven, $T_o$, the initial temperature, $T_c$ of the cake, the radius, $R$, of the cake pan, and the depth, $Z$, of the cake as well as the value of $\mathcal{D}^{\prime}$ into Eq.~\ref{eq2}.  This gives the internal temperature of the cake, $T_I(t)$ for time t, $0 \leq t \leq t_1$.
\item $T_I(t_1)$ becomes  the initial temperature of the  temperature distribution for times t, $ t_1 < t \leq t_2$.  
\item Expand $T_I(t_1)$ in terms of the eigenfunctions, $\Psi_{mj}$, Eq.~\ref{eq9}, to obtain the temperature distribution $T_{II}(t)$  during the time interval $ t_1 < t \leq t_2$.
\item $T_{II}(t_2)$ becomes  the initial temperature of the  temperature distribution for times t, $t_2 < t \leq t_{\text{f}}$.  
\item Expand $T_{II}(t_2)$ in terms of the eigenfunctions, $\Phi_{mn}$, Eq.~\ref{eq2a}, to obtain the temperature distribution $T_{III}(t)$  during the time interval $t_2 < t \leq t_{\text{f}}$.  The diffusion coefficient $\mathcal{D}^{\prime}$ is substituted for $\mathcal{D}$ in the $\Phi_{mn}$.
\end{enumerate}
In practice, the evaluation of  the infinite series for the temperature distributions in Intervals I,II, and III requires that each series be truncated at a finite number of terms, depending on the degree of accuracy desired.  For Interval I, assume that Eq.~\ref{eq2} is terminated for integers $M$, ($m \leq M$) and $N$, ($n \leq N$). In order to achieve approximately the same degree of accuracy in Interval III, we apply the same values to the corresponding infinite series.  For Interval II, the temperature distribution is  expanded using all $\Psi_{mj}$, also, for values of $m \leq M$, but selection of the truncation for $j$ is more complicated. All values of $j$ such that 
\begin{equation}
\lambda_{mj} \leq \frac{(N +1/2)\pi}{Z-a}  \label{eq16}
\end{equation}
are used. We note that solving for all eigenvalues $\lambda_{mj}$  is nontrivial because there is not a  one to one correspondence between those in Interval II and those in Intervals~I and~III. 

\begin{figure}[ht]
\begin{center}
\epsfig{file=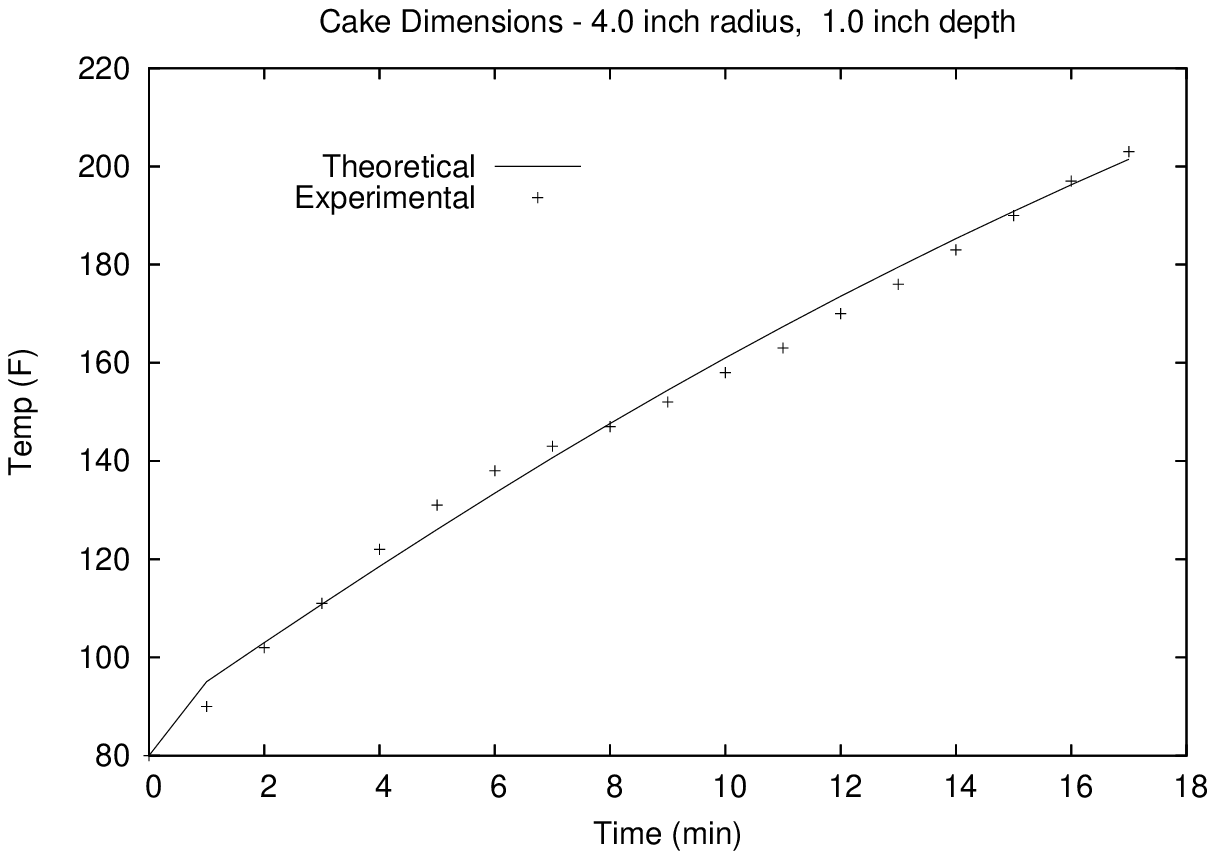,height=6cm} 
\end{center}
\caption{Shown is a theoretical fit to temperature as a function of time at the center of a cake, $4.0\inc$ radius and $1.0\inc$ depth.  The relevant parameters are given in Table~\ref{table2}.}
\label{fig5}
\end{figure}

\begin{figure}
\begin{center}
\epsfig{file=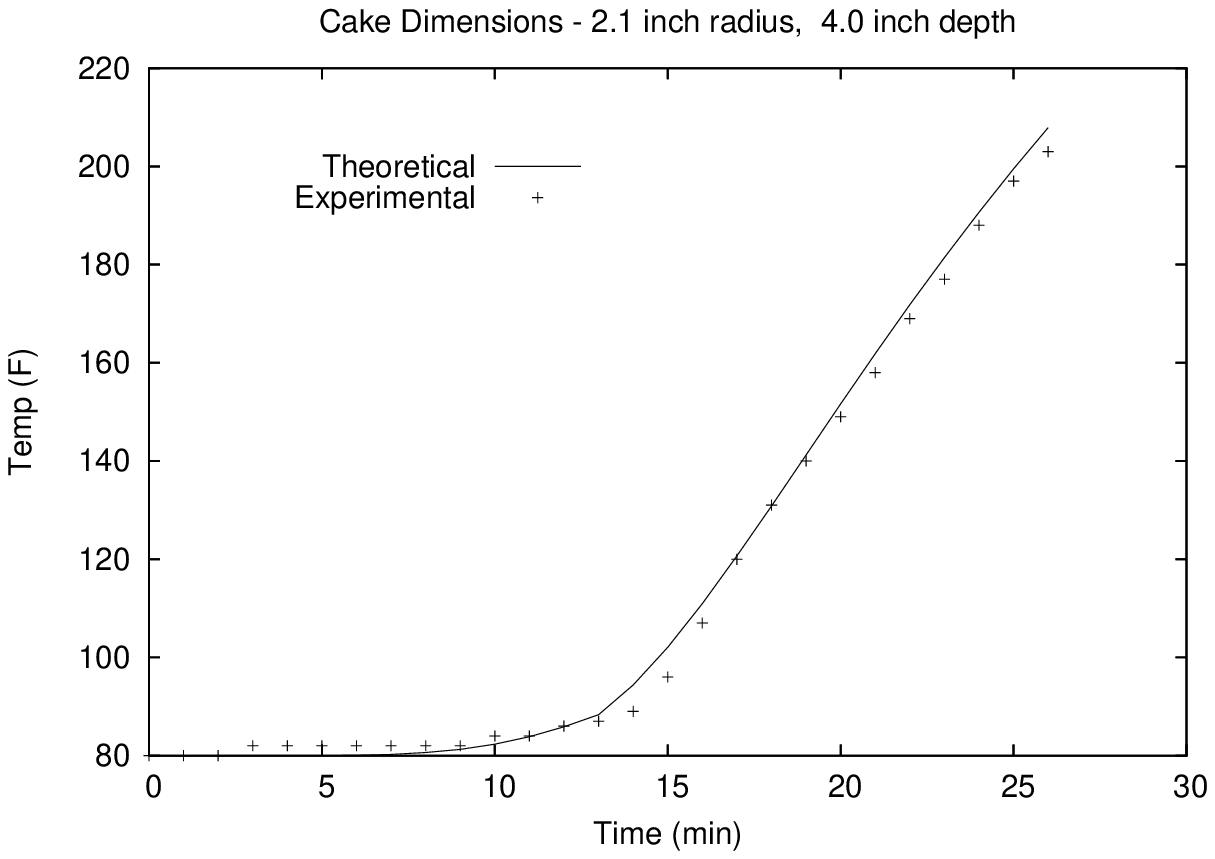,height=6cm} 
\end{center}
\caption{Shown is a theoretical fit to temperature as a function of time at the center of a cake, $2.1\inc$ radius and $4.0\inc$ depth.  The relevant parameters are given in Table~\ref{table2}.}
\label{fig6}
\end{figure}

\begin{figure}
\begin{center}
\epsfig{file=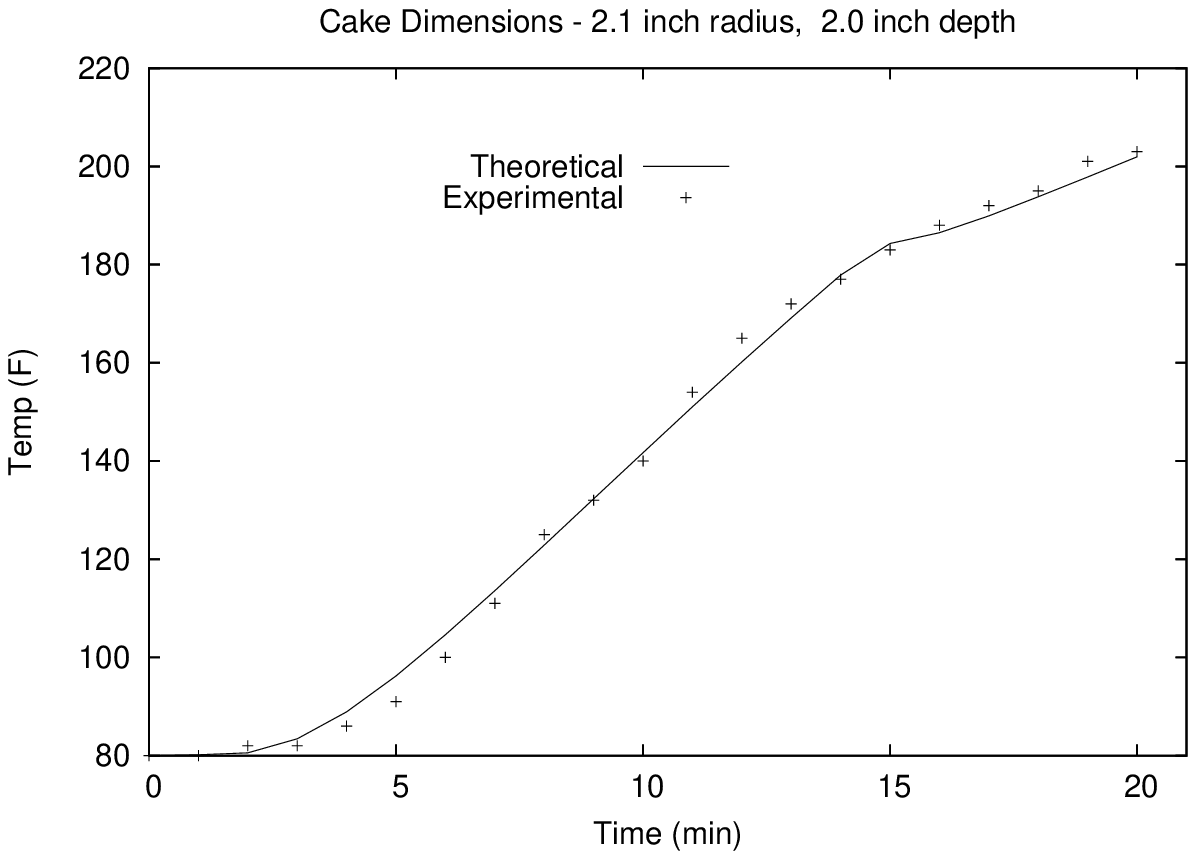,height=6cm} 
\end{center}
\caption{Shown is a theoretical fit to temperature as a function of time at the center of a cake, $2.1\inc$ inch radius and $2.0\inc$ depth.  The relevant parameters are given in Table~\ref{table2}. }
\label{fig7}
\end{figure}

\begin{figure}
\begin{center}
\epsfig{file=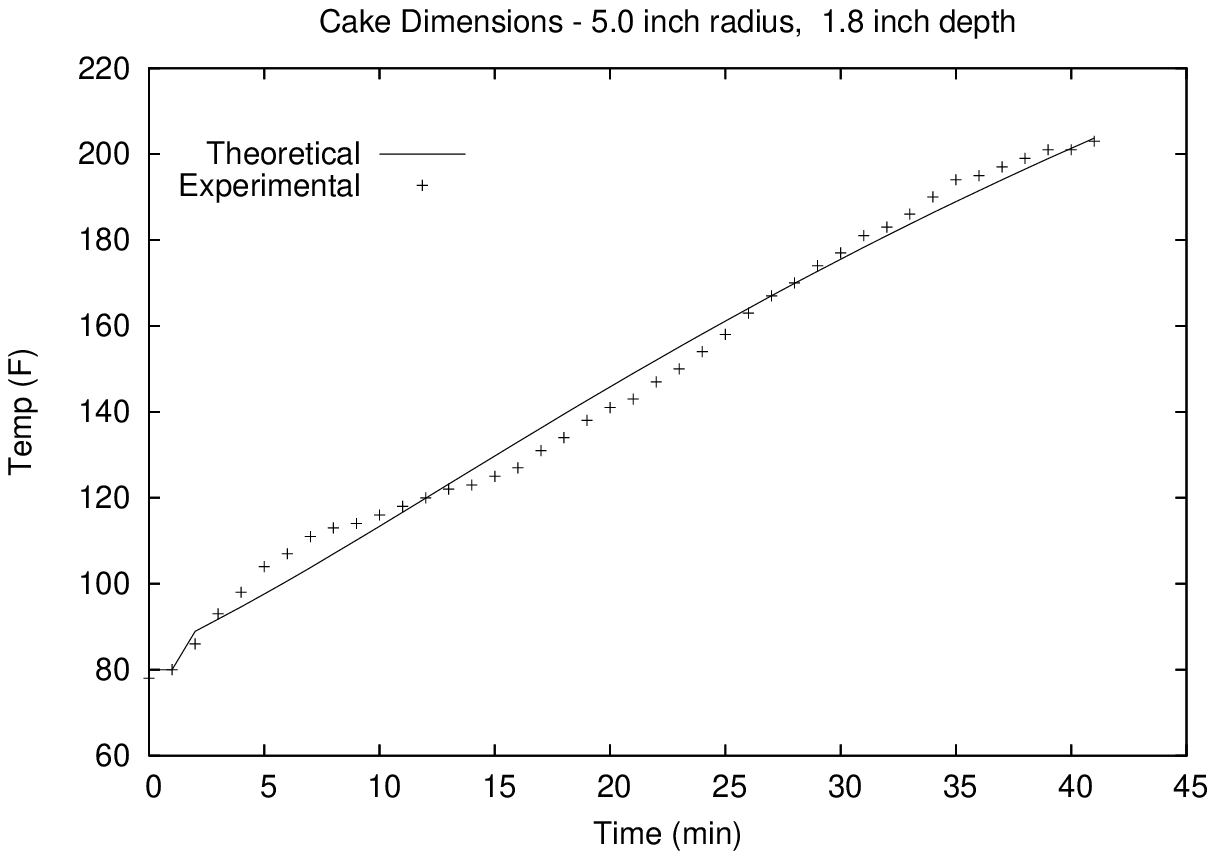,height=6cm} 
\end{center}
\caption{ Shown is a theoretical fit to temperature as a function of time at the center of a cake, $5.0\inc$ radius and  $1.8\inc$ depth.  The relevant parameters are given in Table~\ref{table2}.}
\label{fig8}
\end{figure}

We have used these expansions to model the temperature at the center of the cake for four representative, sets of data.  We summarize the results in Table~\ref{table2} and Figures~\ref{fig5} through~\ref{fig8}.\footnote{Computations   have been performed using Maple software running under Windows XP on a one GHz, dual processor, Pentium III computer.  The optimization of adjustable parameters in the model has been achieved by trial and error, as there has been no simple or straightforward way to interface the model calculations to optimization software.} The data  have been selected to span the range of  cake dimensions.  As is apparent in Figures~\ref{fig5} through~\ref{fig8} it is possible to obtain a reasonable fit to each set of data by varying the four adjustable parameters in the model, $\mathcal{D}$, $\mathcal{D}^\prime$, $t_1$, $t_2$.  The primary purpose of constructing these models has been to provide incite into the baking process so that quantitative estimates can be obtained for the baking time of cakes of various dimensions. These results are consistent with the qualitative aspects of the revised model.  It appears, in accordance with Table~\ref{table2}, that cakes of larger depth, although taking longer to bake, have an effective diffusion coefficient which is greater than  cakes of smaller depth.  On the other hand, cakes of smaller radius seem to have an effective diffusion coefficient which is greater than cakes of larger radius.  

In cakes of less depth moisture can evaporate from the top surface of the cake relatively quickly causing the cake to dry out.  This causes a decrease in the diffusion coefficient because of lack of moisture content. For an amount of baking time, $t$, we hypothesize that the moisture content within the cake scales, approximately, as $s_{tz}$,
\begin{equation}
s_{tz} = \frac{t}{Z^2} \: . \label{eq17a}
\end{equation}
This type of behavior is typical in random walk processes since diffusion times scale as some $\text{( characteristic length)}^2$.

Moisture, also, diffuses from the sides of the cake, inward.  In situations when the radius of the cake is relatively less than its depth, moisture in the interior of the cake may remain constant, or possibly increase, even though moisture is continuing to evaporate from the top cake surface.  For an amount of baking time, $t$, we further hypothesize that  the moisture content scales, approximately, as  $s_{rz}$, 
\begin{equation}
s_{rz} = \frac{t/R^2}{t/Z^2}=\frac{Z^2}{R^2} \: . \label{eq17b}
\end{equation}

We acknowledge that these hypotheses regarding the scaling behavior of the moisture content are highly speculative, and that the degree to which they can be justified rests on how successfully they can be applied to predicting the baking time of cakes of various dimensions.

\begin{table}
\center 
\begin{tabular} {|  c   c  c | c  c  c  c | c  c |  } \hline
 \multicolumn{3}{ | c |}{Data} & \multicolumn{4}{ c |}{Adjustable Parameters} & \multicolumn{2}{ c |}{Accuracy} \\ \hline
   Radius (\text{in}) & Depth (\text{in})   & $t_{\text{f}}$ (min) & $\mathcal{D} (\text{in}^2/\text{min})$ & $\mathcal{D}^{\prime} (\text{in}^2/\text{min})$ & $t_1$ (min)  & $t_2$ (min) & M & N \\ \hline 
    4.0  &  1.0 & 17 & .029 &  .0035   & 2 & 2 &  10 & 10 \\ \hline
  2.1 & 4.0 &  26  & .012 & .047   & 10  & $\infty$ &  10 & 10  \\ \hline
 2.1  & 2.0 &  20  & .025 &  .010  & 1  & 15 &  10 & 10  \\ \hline
  5.0 & 1.8 & 41    & .035 & .005  & 2 & 2 &  10 & 10 \\ \hline
6.5  & 1.6 &  35  & -- &  --  & --  & -- &  -- & --  \\ \hline
  5.0 & 1.0 & 20    & -- & --  & -- & -- &  -- & -- \\ \hline
  2.1 & 1.8 & 19    & -- & --  & -- & -- &  -- & -- \\ \hline
\end{tabular}
\caption{Shown is a summary of the data used in this study. A representative sample of these data has been selected for purposes of modeling. The values  $M$ and $N$ have been selected to insure that the theoretical curves are relatively smooth. The adjustable parameters have been determined by trial and error.}
\label{table2}
\end{table}

\setcounter{equation}{0}

\section{Estimating the Baking Time of a Cake}
\label{sect4}
We now utilize the revised model of  Section~\ref{sect3_3} to obtain an explicit formula for  the baking time of a cake.\footnote{In this Section the diameter, rather than the radius, is used in all formulae. } 
In the revised model we have assumed that as a cake bakes its  diffusion coefficient  changes, primarily, because of evaporation of moisture at the top surface of the cake; however, for a given cake we can calculate, using Eq.~\ref{eq4}, an effective diffusion coefficient, $\mathcal{D}_{\text{eff}}$,     which is constant during the baking process.   Furthermore, in accord with the revised model the functional form of $\mathcal{D}_{\text{eff}}$ should be consistent with Eqs.~\ref{eq17a} and~\ref{eq17b}. If the functional form  is known,  it provides a relationship between the baking time and physical dimensions of the cake, making it possible to solve for the baking time  as a function of the cake dimensions. 

For each of the seven data entries in Table~\ref{table2}, we estimate an effective  diffusion coefficient using Eq.~\ref{eq4}.  We, then, fit to these data an expression for  $\mathcal{D}_{\text{eff}}$ of the following     form,
\begin{equation}
  \mathcal{D}_{\text{eff}} = a_{0} - a_{1} \exp \left[ - a_{2} \:  \left( \frac{ Z^{2} }{ t_f } \right) \left( \frac{Z^2}{D^2} \right) \right]  \:. \label{eq18}
\end{equation}
 The adjustable parameters are estimated to be $a_0 = 2.57 \times 10^{-2}$, $a_1 = 2.07 \times 10^{-2}$, and $a_2 = 19.2$.  
The functional form of Eq.~\ref{eq18} has been chosen primarly because it can be   interpreted in a relatively straightforward and physically intuitive manner.  We have tried other formulae with more physical appeal, i.e.\@ different dependencies on the baking time and cake dimensions; however, when inverted to  estimate baking times they yield multiple roots for a cake of specified dimensions.  Generally, these additional roots have no obvious physical interpretation.   We have also varied the functional form of Eq.~\ref{eq18} among those which are not plagued with multiple roots for the baking time and have found the calculated baking times not to be drastically different from those obtained using Eq.~\ref{eq18}.

\begin{figure}
\begin{center}
\epsfig{file=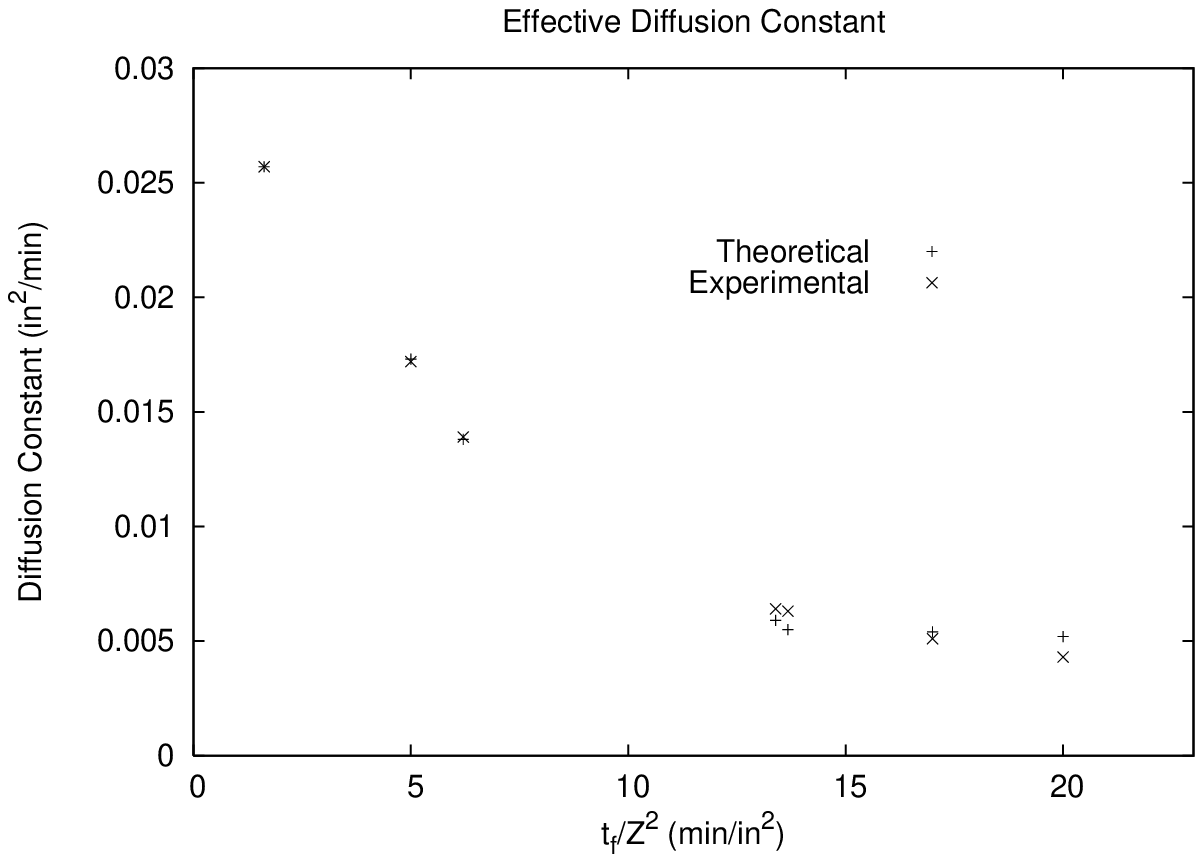,height=6cm} 
\end{center}
\caption{Shown are the experimental values of the effective diffusion coefficient and the theoretical fit to these values.  The independent variable has been selected based on the discussion of Section~\ref{sect3_3}. }
\label{fig12}
\end{figure}

\begin{figure}
\begin{center}
\epsfig{file=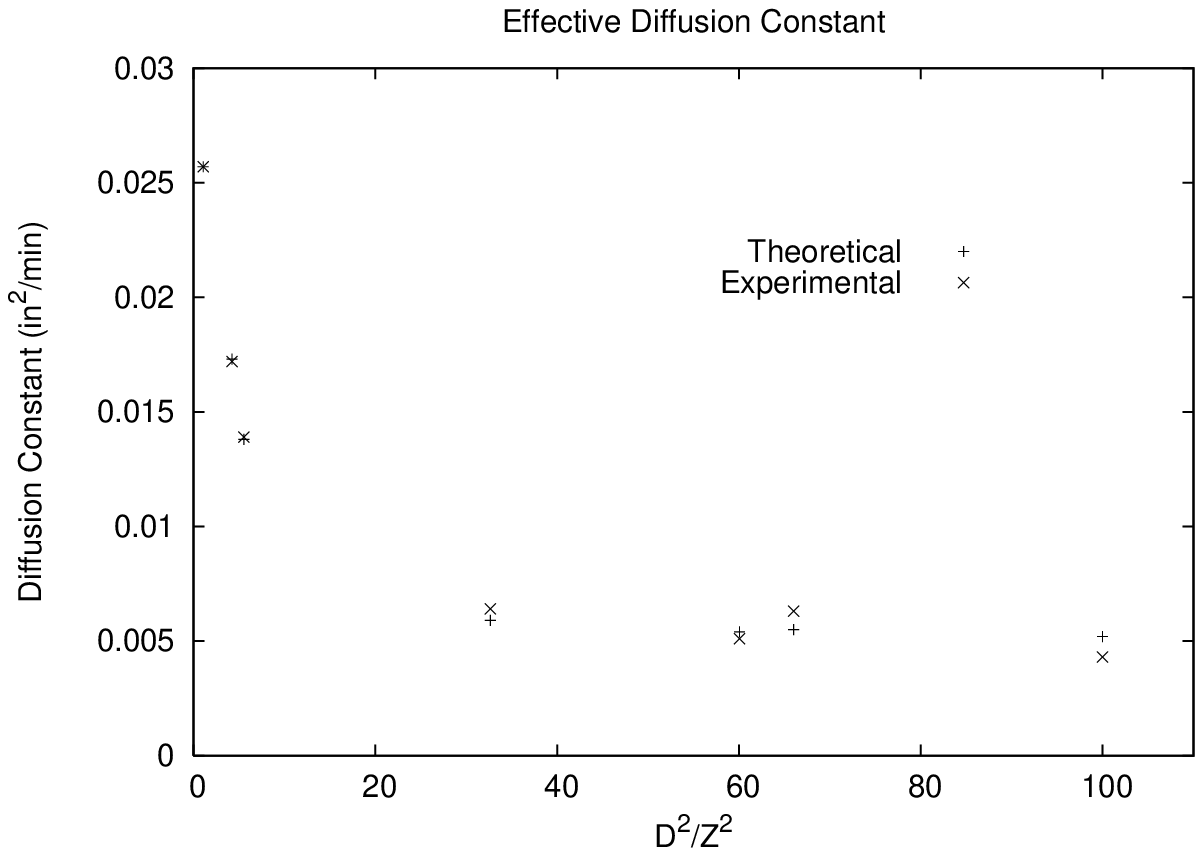,height=6cm} 
\end{center}
\caption{ Shown are the experimental values of the effective diffusion coefficient and the theoretical fit to these values.  The independent variable has been selected based on the discussion of Section~\ref{sect3_3}.}
\label{fig13}
\end{figure}

The effective diffusion coefficient depends explicitly on the baking time, the diameter of the pan, and depth of the cake batter.  In Figures~\ref{fig12} and~\ref{fig13} we show the experimental values of $\mathcal{D}_{\text{eff}}$  and the theoretical fit.  Rather than plotting them as a function of the baking time, diameter of the pan, and depth of the batter, we have plotted them as a function $t_f/Z^2$ and $D^2/Z^2$.
The adjustable parameters in Eq.~\ref{eq18} have a simple, physical  interpretation.  If values of $t_f$, $D$, or $Z$ are such that the exponential becomes vanishingly small, the cake batter should be maximally moist and  $\mathcal{D}_{\text{eff}}$   is $a_0$.  On the other hand, if their values are such that  the exponential approaches one, the cake should be totally dry with the effective diffusion coefficient being $a_{0}-a_{1}$.  The parameter $a_2$ relates to how readily  moisture evaporates from the cake.   These three parameters should also depend on the initial moisture content of the cake batter.

Combining Eqs.~\ref{eq4} and~\ref{eq18} we obtain an implicit formula for the baking time  as a function of depth, diameter, oven temperature,  initial temperature of the cake batter, and final temperature of the cake:  
\begin{equation}
   K = - \mathcal{D}_{\text{eff}} \; t_f    
\begin{cases}
\frac{4 x_{1,0}^{2}}{D^{2}} + \frac{\pi^{2}}{Z^{2}}  
&  \\    \text{\hspace{.5in} if $C_1 \leq 1$ and $C_2 \leq 1$}  \\ 
 \frac{\pi^{2}}{Z^{2}}  
& \\ \text{\hspace{.5in} if $C_1>1$}  \hspace{.5in}        \\ 
\frac{4 x_{1,0}^{2}}{D^{2}}   
& \\  \text{\hspace{.5in}  if $C_2 >1$} \end{cases}
\: ,   \label{eq19}
\end{equation}
where $K$ is obtained from 
 \begin{equation}
 \exp(-K) = \left( \frac{T_{f}-T_{c}}{T_{o}-T_{c}} \right)
\begin{cases}
\frac{ \pi \;x_{1,0} \; J_1(x_{1,0})}{8}     
&  \\    \text{\hspace{.5in} if $C_1 \leq 1$ and $C_2 \leq 1$}  \\ \\   
\frac{ \pi}{4}  
 & \\ \text{\hspace{.5in} if $C_1>1$}  \hspace{.5in}        \\ \\ 
\frac{  \;x_{1,0} \; J_1(x_{1,0})}{2} 
  & \\  \text{\hspace{.5in}  if $C_2 >1$}   
\end{cases}
\: .   \label{eq20}
\end{equation}
The conditions $C_1$ and $C_2$ are defined by Eqs.~\ref{eq6} and~\ref{eq6}$^\prime$.

\begin{figure}
\begin{center}
\epsfig{file=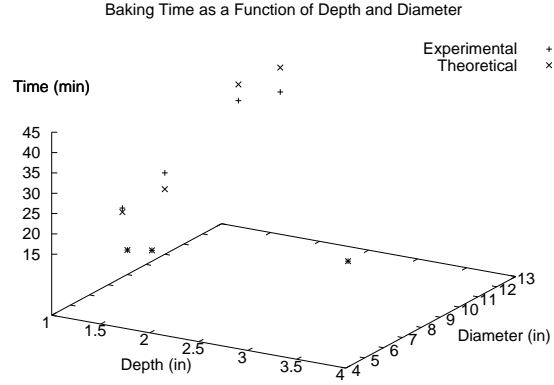,height=6cm} 
\end{center}
\caption{The theoretical and measured baking times of cakes of various depths and diameters are shown. }
\label{fig14}
\end{figure}

We use Eqs.~\ref{eq19} and~\ref{eq20} to estimate baking times of cakes of various dimensions used in this study and compare them to the actual baking times.  These results are presented in Figure~\ref{fig14} and Table~\ref{table3}. It should be noted that Eqs.~\ref{eq19} and~\ref{eq20} exhibit the scaling behavior that is typical of diffusion processes, i.e.\@ baking times are proportional to $\text{(characteristic length)}^2$.

\begin{table}
\center 
\begin{tabular} {|c c c c c|  } \hline
  $t_{\text{f}}$ (min) & $t_{\text{theor}}$ (min) & $t_{\text{linreg}}$ (min)&Diameter (\text{in}) & Depth (\text{in})   \\ \hline 
17 &  16 & 21 & 8.0  &  1.0   \\ \hline
26 & 26 & 29 & 4.1 & 4.0     \\ \hline
20& 20 & 18 & 4.1  & 2.0    \\ \hline
41 & 45 & 30 &  9.9 & 1.8   \\ \hline
35 & 41 & 37 & 13.0  & 1.6    \\ \hline
20 & 16 & 25 & 9.9 & 1.0   \\ \hline
19 & 19 & 17 & 4.1 & 1.8  \\ \hline
\end{tabular}
\caption{Baking times and those predicted by the theoretical model are presented for the  diameters and the depths of cakes used in this study. For comparison the baking time predicted by linear regression is also presented.}
\label{table3}
\end{table}

Philosophically speaking, a semi-empirical model should accomplish more than regurgitate what is already known:  
\begin{enumerate}
\item it should have predictive ability;
\item it should also be valid in that it is consistent with established physical principles;  
\item it should not have an inordinately large number of freely adjustable parameters. 
\end{enumerate}
Consider the four hypothetical cakes whose dimensions and predicted baking times are given in Table~\ref{table4}  For various reasons these cakes are impractical to bake.  Nonetheless, they do provide an arena for qualitatively exploring the theoretical basis of Eqs.~\ref{eq19} and~\ref{eq20}.  Even to someone with little cooking experience it is not surprising that cake number two requires 17 minutes to bake, given that a cake $8\inc$ diameter and $1\inc$ depth requires 17 minutes baking time (See Table~\ref{table3}).  It may be surprising that cake number three requires so little time to bake, given how long it takes cake number two to bake.  The reason is that the moisture content of cake number three remains high because very little evaporation takes place during the cooking process.  As a result, the diffusion coefficient remains relatively large so that the cake bakes quickly.  Cake number four requires nearly 24 hours to bake.  It is interesting to note that the baking time of cake number four reflects the scaling property of diffusion processes, i.e.\@ its physical dimensions are thirty times those of cake number one and is therefore predicted to require 900 times as long to bake.

The semi-empirical model for the baking time has three adjustable parameters. From a practical perspective one may ask if performing a linear regression, which also has three adjustable parameters, would produce a model with as good or possibly better predictive ability than the model presented.  The result of a linearly regressing  baking time on diameter and depth is
\begin{equation}
t_f = -3.00 + 2.34 D +5.74 Z \: .  \label{eq21}
\end{equation} 
In Table~\ref{table3} we present the baking times predicted by Eq.~\ref{eq21} for comparison to those predicted according to theory.
 
Although, strictly speaking, our analysis has been applied to the g\'enoise we can apply the results, qualitatively, to baking in general.  The difference between cake batter and pastry dough is not so much a matter of different ingredients, but rather the proportions of the same ingredients.  Both typically are composed of flour, fats, and ingredients which contain water, such as  eggs, milk, juices, etc. Typically, pastry dough contains a very small amount of liquid in proportion to other ingredients.  This explains why it takes such a long time, relatively speaking , to pre-bake a pastry shell (15 to 30 minutes, for example), even though, pastry shells are generally, not much more than $\frac{1}{8}\inc$ to   $\frac{1}{4}\inc$ in depth.  Cookie dough which lies  between pastry dough  and cake batter, with respect to moisture content, also requires a relatively longer time to bake than cake batter.  Cookies require typically between ten and fifteen minutes to bake, even though they are usually not more than $\frac{1}{2}\inc$ thick.

\begin{table}
\center 
\begin{tabular} {|c c c c |  } \hline
 Cake  &  $t_{\text{theor}}$ (min) & Diameter (\text{in}) & Depth (\text{in})   \\ \hline 
 1 & 1.56  & 1.00  &  1.00   \\ \hline
2 & 17.0 &  30.0 & 1.00     \\ \hline
3 & 1.82 & 1.00  & 30.0    \\ \hline
4 & $1.40 \times 10^3$ &  30.0 & 30.0   \\ \hline
\end{tabular}
\caption{The predicted baking times of four hypothetical cakes are presented.}
\label{table4}
\end{table}

\setcounter{equation}{0}

\section{An Irrelevant but Intriguing Digression}
\label{sect5}
The similarity between the Diffusion equation and the Schr\"{o}dinger equation is striking.  We replace $t$ in the Diffusion equation by $it/\hbar$ to obtain the Schr\"{o}dinger equation.  The duality between solutions of the Diffusion equation and Schr\"{o}dinger equation is typically not insightful since the corresponding solutions of the  Schr\"{o}dinger equation are those in the absence of a potential.  Consider, however, the eigenfunctions, Eq.~\ref{eq9}.  They are dual to the eigenfunctions of a particle confined to a cylindrical box.  What is atypical is the mass of the particle assumes different values in the regions $0 \leq z \leq a$ and $ a < z \leq Z$. The mass, $\mu$, is related to the diffusion coefficient, $\mathcal{D}$, $\mu=\hbar^2/(2\mathcal{D})$.  It is  common for situations to exist where the  diffusion coefficient  varies with position; however, it is less common for the mass of a particle to vary with position.  Nonetheless, there do exist situations where the mass of a particle, or quantum mechanical system, may be different in different regions of space.
         Quantum systems with spatially dependent effective mass have been used, extensively, in modeling  the physical properties of various microstructures and semiconductor interfaces in condensed matter.\cite{alhaidari03} Such specialized models have been applied to studying the electronic properties of semiconductors~\cite{bastard88}, quantum wells and quantum dots~\cite{harrison00},\cite{serra97}, $^{3}$He clusters~\cite{barranco97}, quantum clusters~\cite{desaavedra94}, and graded alloys and heterostructures.~\cite{einevoll90},\cite{einevoll90b},\cite{weisbuch90}.   

\begin{figure}[ht]
\begin{center}
\input{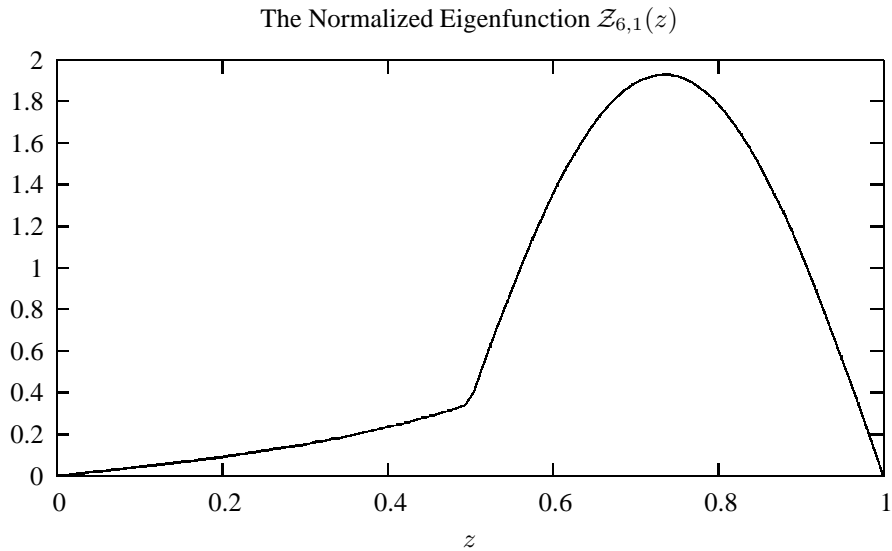} 
\end{center}
\caption{The eigenfunction $ \mathcal{Z}_{6,1}$  exhibits both decaying and oscillatory behavior in regions $0 \leq z\leq 1/2$ and $1/2 < z \leq 1$, respectively.  In addition, $\frac{\partial \mathcal{Z}_{6,1}}{\partial z} $ is discontinuous at $z = a$, where $a = 1/2$.}
\label{fig9}
\end{figure}

If one is guided by the seeming duality between the Diffusion equation and Schr\"{o}dinger equation, then the Schr\"{o}dinger equation is  
\begin{equation}
 - \nabla \left( \frac{\hbar^2}{2\mu} \right) \nabla \psi  = -\frac{\hbar}{i}\frac{\partial \psi}{\partial t}   \: ,  \label{eq22}
\end{equation}
This insures  that the Schr\"{o}dinger equation conserves probability so that   the continuity equation remains true,
\begin{equation}
\nabla \cdot \mathbf{J} = -\frac{\partial P}{\partial t} \: , \label{eq23}
\end{equation}
where the current density, $\mathbf{J}$, is given by
\begin{equation} 
\mathbf{J} = \frac{\hbar}{2\mu i} \left[ \psi^\ast \nabla \psi   -   \left( \nabla \psi^\ast \right)\psi \right] \: , \label{eq24}
\end{equation}
and the probability density, $P$, is given by
\begin{equation} 
P= \psi^\ast \psi    \: .  \label{eq25}
\end{equation}
If $\mu$ is placed to the left or right of the derivatives, except for specific combinations of such, there will be sources/sinks of probability, as can be shown by direct calculation.  A consequence which results from the unusual placement of $\mu$ in Eq.~\ref{eq22} is that when the mass, $\mu$,  discontinuously changes values between two regions,  the spatial derivative of the wavefunction is discontinuous at the interface between the two regions.  This effect is apparent in Eq.~\ref{eq13} whenever $ \mathcal{D} \neq\mathcal{D}^\prime $.  This effect is demonstrated in Figures~\ref{fig9},~\ref{fig10}, and~\ref{fig11}, which are examples of  the $z$-dependent part of the eigenfunctions of the Diffusion equation given in Eq.~\ref{eq10}. 

Another apparent effect which is evident from a detailed analysis of Eq.~\ref{eq15} is that for sufficiently large values of $m$ and sufficiently small values of $j$ there are solutions with $\lambda^2_{mj} < 0 $.  The implication is that some of the eigenfunctions are hybrid in that for $ 0 \leq z \leq a$ they decay and for $ a < z \leq Z$ they oscillate.   The corresponding  eigenfunctions of the Schr\"{o}dinger equation  also exhibit this same behavior,  decaying in the region of smaller mass and oscillating in the region of larger mass.  This effect is shown in Figures~\ref{fig9},~\ref{fig10}, and~\ref{fig11}.  
In Figure~\ref{fig11} the eigenfunction, whose value of $j$ is  the largest of three examples, is  oscillatory in both regions.  This effect can also be shown to persist under more general boundary conditions than those imposed by the cake pan.\footnote{The effect seems only to be a consequence of the fact that the mass  assumes different values in  two contiguous spatial regions.} From the quantum mechanical perspective the values of $m$ are associated with momentum tangent to the interface separating the regions of different values of  mass.  Thus, if the momentum tangent to the interface is sufficiently large,  the wave function damps out in the region of less mass.  This is reminiscent of the evanescent modes associated when light propagates along a wave guide. 
\begin{figure}[ht]
\begin{center}
\input{teigen62} 
\end{center}
\caption{ The eigenfunction, $ \mathcal{Z}_{6,2}$ exhibits both decaying and oscillatory behavior in regions $0 \leq z \leq 1/2$ and $1/2 < z\leq 1$, respectively. In addition, $\frac{\partial \mathcal{Z}_{6,2}}{\partial z} $ is discontinuous at $z = a$, where $a = 1/2$. }
\label{fig10}
\end{figure}
\begin{figure}[ht]
\begin{center}
\input{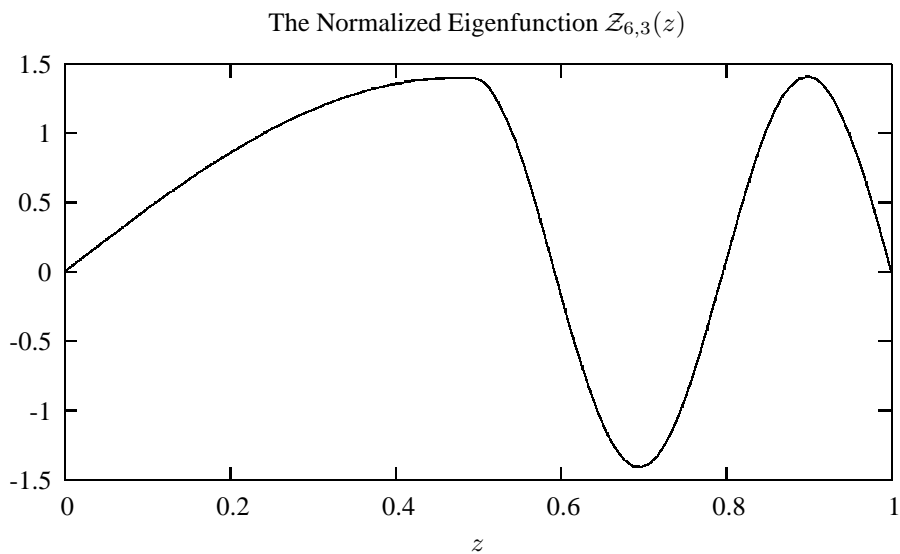} 
\end{center}
\caption{ The eigenfunction, $ \mathcal{Z}_{6,3}$  exhibits only oscillatory behavior in the region $0 \leq z\leq 1$. In addition, $\frac{\partial \mathcal{Z}_{6,3}}{\partial z} $ is discontinuous at $z = a$, where $a = 1/2$.}
\label{fig11}
\end{figure}

If one considers the Schr\"{o}dinger equation as the non-relativstic limit of the Klein-Gordon equation, then one should also include the term $mc^2$ on the left side of Eq.~\ref{eq22}. It can, then, be shown that the hybrid eigenfunctions are no longer solutions.

\setcounter{equation}{0}
\section{Conclusions}
\label{sect6}
The primary emphasis of this study has been to explain how modifying a cake recipe by changing either the dimensions of the cake  or the amount of cake batter alters the baking time.  Our analysis has been restricted to one particular cake, the g\'{e}noise. Based on data collected and ensuing analysis we have concluded that conduction is the primary mechanism of heat transfer and that the Diffusion equation provides a theoretical framework adequate for describing the baking process; however, the diffusion coefficient does not remain constant during the baking process. From detailed analysis of the data  we have concluded that the diffusion coefficient changes during the baking process, principally, because of  evaporation of moisture at the top surface of the cake. Consequently, we have approximated the cake baking process  as one in which the diffusion coefficient, effectively, assumes a constant value during the baking process.   Its value depends on the diameter, depth, and baking time of the cake, all of which affect the moisture  content of the cake. We, then, have proposed a semi-empirical formula for the effective diffusion coefficient consistent with the physical principles underlying the model.  The formula has three adjustable parameters whose values have been estimated from the data.  We have inverted the formula to obtain the baking time of a cake as a function of diameter and depth.  The resulting formula exhibits scaling behavior typical of diffusion processes: the baking time scales as the $(\text{the characteristic size })^2$ of the cake.  Although we have not solved the problem of how long is required to bake a cake, in general, we have offered a qualitative explanation of those factors which most importantly determine baking times.   In addition to the dimensions of the cake we have proposed that the moisture content of a cake is the dominant factor affecting its baking time.  This proposition explains why  pastry dough and cookie dough  require such a relatively long time to bake in comparison to cake batter,  after  accounting for differences in their physical dimensions.  

As part of the analysis and modeling process   we have required solutions of the Diffusion equation for two cylindrical media of the same diameter, separated by a common circular boundary.  The diffusion coefficient assumes a different constant value in each medium. These solutions are interpreted naturally and straightforwardly in the context of the Diffusion equation; however, when interpreted in the context of the Schr\"{o}dinger equation, they are somewhat peculiar.  The solutions describe a system whose mass assumes different values in two different regions of space.  Furthermore, the solutions exhibit characteristics similar to the evanescent modes associated with light waves propagating in a wave guide; however, when we consider the Schr\"{o}dinger equation as a non-relativistic limit of the Klein-Gordon equation so that it includes a mass term, these are no longer solutions.

\newpage
\bibliography{cake}  
\end{document}